\documentclass [aps,prl,twocolumn,nofootinbib,superscriptaddress]{revtex4}
\usepackage{epsfig}
\usepackage{graphicx}
\usepackage{amsmath}
\usepackage{amssymb}
\usepackage{bm}
\allowdisplaybreaks

\begin{document}


\boldmath
\title{Constraints on Nonrelativistic-QCD Long-Distance Matrix Elements\\
from $J/\psi$ Plus $W$/$Z$ Production at the LHC}
\unboldmath
\author{Mathias Butenschoen}
\affiliation{{II.} Institut f\"ur Theoretische Physik, Universit\"at Hamburg,
Luruper Chaussee 149, 22761 Hamburg, Germany}
\author{Bernd A. Kniehl}
\affiliation{{II.} Institut f\"ur Theoretische Physik, Universit\"at Hamburg,
Luruper Chaussee 149, 22761 Hamburg, Germany}
\date{\today}
\begin{abstract}
  We study the associated production of prompt $J/\psi$ mesons and $W$ or $Z$
  bosons within the factorization approach of nonrelativistic QCD (NRQCD) at
  next-to-leading order in $\alpha_s$, via intermediate color singlet
  ${^3}S_1^{[1]}$ and ${^3}P_J^{[1]}$ and color octet ${^1S}_0^{[8]}$,
  ${^3S}_1^{[8]}$ and ${^3P}_J^{[8]}$ states.
  Requiring for our predictions to be compatible with recent ATLAS measurements
  yields stringent new constraints on charmonium long-distance matrix elements
  (LDMEs) being nonperturbative, process-independent input parameters.
  Considering four popular LDME sets fitted to data of single $J/\psi$
  inclusive production, we find that one is marginally compatible with the
  data, with central predictions typically falling short by a factor of three,
  one is unfavored, the factor of shortfall being about one order of magnitude,
  and two violate cross section positivity for direct $J/\psi+W/Z$ production.
  The large rate of prompt $J/\psi$ plus $W$ production observed by ATLAS
  provides strong evidence for the color octet mechanism inherent to NRQCD
  factorization, the leading color singlet contribution entering only at
  $\mathcal{O}(G_F\alpha_s^4)$, beyond the order considered here. 
\end{abstract}

\maketitle

Although heavy quarkonia have been discovered already in 1974, the underlying mechanisms governing their production in high-energy collisions are still not fully understood. The most prominent approach is via the factorization theorem of nonrelativistic QCD (NRQCD)~\cite{Caswell:1985ui,Bodwin:1994jh}. According to it, the production cross section of quarkonium $H$ factorizes into perturbative short-distance cross sections of heavy quark-antiquark bound-state production and supposedly universal nonperturbative long-distance matrix elements (LDMEs) $\langle {\cal O}^H (n) \rangle$, where $n={}^{2S+1}L_J^{[1,8]}$ denotes the quarkonic Fock state, in color singlet ``$[1]$'' or octet ``$[8]$'' configuration. Velocity ($v$) scaling rules \cite{Lepage:1992tx} impose a strong hierarchy on the $\langle {\cal O}^H (n) \rangle$ values, leading to a double expansion in the strong-coupling constant $\alpha_s$ and $v$. For $H=J/\psi$ and $\psi(2S)$, the LDMEs of $n={^3S}_1^{[1]}$ are leading in $v$ and those of $n={^1S}_0^{[8]}$, ${^3S}_1^{[8]}$ and ${^3P}_J^{[8]}$ are subleading. For $H=\chi_{cJ}$, the LDMEs of $n={^3P}_J^{[1]}$ and ${^3}S_1^{[8]}$ are both leading in $v$.

The available charmonium LDME sets have all been extracted from data of
single inclusive production.
Thanks to the high luminosity meanwhile achieved by the LHC, also double
production and associated production with bottomonia, $W$, and $Z$ bosons have
been studied there, which can inject orthogonal information into
LDME determinations.
The goal of this letter is to provide the first complete analysis of
prompt-$J/\psi$ plus $W$ or $Z$ hadroproduction at next-to-leading order (NLO)
in $\alpha_s$, i.e.\ through $\mathcal{O}(G_F \alpha_s^3)$.
Invoking QCD and NRQCD factorization, we calculate the cross sections as
\begin{align}
 &\sigma(pp\to J/\psi+W/Z+X) = \sum_H \mathrm{Br} (H\to J/\psi) \nonumber \\
  &\quad \times \sum_n \tilde{\sigma}(pp\to c\overline{c}[n]+W/Z+X) \langle {\cal O}^H (n) \rangle, \\
 &\tilde{\sigma}(pp\to c\overline{c}[n]+W/Z+X) = \sum_{a,b} \int dx_a dx_b  \nonumber \\  
 &\quad \times f_{a/p}(x_a) f_{b/p}(x_b)\hat{\sigma}(ab\to c\overline{c}[n]+W/Z+X),\label{eq:xs}
\end{align}
where $H=J/\psi$, $\psi(2S)$, and $\chi_{cJ}$, with $J=0,1,2$, and $n$ runs over all Fock states specified above. $\hat{\sigma}(ab\to c\overline{c}[n]+W/Z+X)$ are the partonic cross
sections, evaluated as perturbative expansions in $\alpha_s$;
$f_{a/p}(x)$ is the parton density function (PDF) of parton $a$ in the proton;
$a,b$ include the up, down, strange (anti)quarks, and the gluon;
$\mathrm{Br}(H\to J/\psi)$ are the decay branching fractions, including
$\mathrm{Br}(J/\psi\to J/\psi)=1$ for ease of notation. 
In the $W$ case, where $W^\pm$ is summed over, only $n={^3}S_1^{[8]}$ contributes at leading order (LO) in $\alpha_s$.

Partial results may be found in the literature.
The LO results have already been obtained two decades ago \cite{Kniehl:2002wd}.
At NLO, the ${^1}S_0^{[8]}$, ${^3}S_1^{[8]}$, and ${^3}P_J^{[8]}$ channels have been
considered in the $W$ case \cite{Li:2010hc}, and the 
${^3}S_1^{[1]}$ \cite{Mao:2011kf,Gong:2012ah} and ${^3}S_1^{[8]}$
\cite{Mao:2011kf} channels in the $Z$ case.
We can reproduce these results, with noticeable differences only in Fig.~4 of
Ref.~\cite{Li:2010hc} for the tree-level $c\overline{c}[{^3}P_J^{[8]}]+W$
channel, and fill all gaps by providing the NLO results for the ${^3}P_J^{[1]}$
channels in the $W$ case and the ${^1S}_0^{[8]}$, ${^3P}_J^{[8]}$, and
${^3P}_J^{[1]}$ channels in the $Z$ case.
Unlike the preceding works, we have to consider virtual corrections to $P$-wave
state production.
Albeit $P$-wave state virtual corrections have been tackled for single inclusive
production, the additional $W/Z$ mass scale elevates the complexity of this NLO
NRQCD calculation to an unprecedented level.

Let us now review the main technical aspects of our calculation, starting
with the treatment of $\gamma_5$, which appears in the $W$ and $Z$ axial-vector
couplings and in the spin projection onto the ${^1S}_0^{[8]}$ state.
Adopting the standard scheme \cite{tHooft:1972tcz,Breitenlohner:1977hr,Larin:1993tq,Collins:1984xc}, we use the axial-vector coupling $\gamma_\mu \gamma_5$ in
its antisymmetric form $\frac{1}{2}(\gamma_\mu \gamma_5 - \gamma_5 \gamma_\mu)$, directly replace $\gamma_5$ by $(i/4!) \epsilon_{\mu\nu\rho\sigma} \gamma^\mu \gamma^\nu \gamma^\rho \gamma^\sigma$, employ the relation
\begin{equation}
\epsilon_{\mu_1\mu_2\mu_3\mu_4} \epsilon_{\nu_1\nu_2\nu_3\nu_4} = - \det \left( g_{\mu_i\nu_j} \right),
\label{eq:epsilons}
\end{equation}
and apply the finite axial-vector coupling renormalization (see, e.g., 
Ref.~\cite{Collins:1984xc}).
We explicitly verify that the final results are then independent on whether
we choose a $D$- or four-dimensional metric $g$ in Eq.~(\ref{eq:epsilons}).

We generate, treat and square the amplitudes using FeynArts~\cite{Hahn:2000kx} and custom FORM~\cite{Vermaseren:2000nd} and Mathematica codes.
We reduce the virtual loop integrals to a common set of master integrals using two methods. In the first one, we directly apply integration-by-parts relations generated with AIR~\cite{Anastasiou:2004vj}, while in the second one, we first invoke a custom Passarino-Veltman-type \cite{Passarino:1978jh} tensor reduction, generalized for the case of arbitrary propagator powers and linearly dependent propagator momenta.
We analytically check the agreement of both methods.
As for the master integrals, we implement our own analytic expressions in combination with QCDLoop~\cite{Ellis:2007qk}, checking everything against OneLoop~\cite{vanHameren:2010cp}.
We analytically simplify the resulting expressions and translate them into FORTRAN routines ready for numerical
integration by our custom parallelized version of VEGAS~\cite{Lepage:1977sw}.
We analytically check the ultraviolet and infrared finiteness of our results
and numerically compare our real corrections, after imposing infrared
cutoffs, against HELACOnia~\cite{Shao:2015vga} output.

We organize the phase space integrations using the dipole subtraction procedure outlined in Ref.~\cite{Butenschoen:2019lef}, changing only the momentum mapping of dipole term $V_{3,j}$ (into MapPW6($p_j$,$p_2$)) to cope with the presence of the massive non-QCD particle in the final state. We numerically check that all dipoles reproduce the real corrections in their respective limits and perform the check on the integrated dipoles outlined in
section 4.3 of Ref.~\cite{Butenschoen:2020mzi}.
As a further check, we also implement the phase space slicing procedure along section 3 of Ref.~\cite{Butenschoen:2020mzi} to find numerical agreement.
We recover the notion \cite{Butenschoen:2020mzi} that dipole subtraction
significantly outperforms phase space slicing as for precision and speed.

We renormalize the charm quark mass in the on-shell scheme to be $m_c=1.5$~GeV
and take the charmonia to have mass $2m_c$ for definiteness.
We express all electroweak couplings in terms of Fermi's constant $G_F$ and the
on-shell $W$ and $Z$ boson masses $M_W$ and $M_Z$.
We adopt from Ref.~\cite{Zyla:2020zbs} the values
$G_F=1.1664\times 10^{-5}$~GeV$^{-2}$, $M_W=80.379$~GeV, $M_Z=91.188$~GeV,
$|V_\mathrm{ud}|=0.9737$, $|V_\mathrm{us}|=0.2245$, and all relevant branching
fractions.
At LO (NLO), we use the CTEQ6L1 (CTEQ6M) proton PDFs \cite{Pumplin:2002vw}
with asymptotic scale parameter $\Lambda_\mathrm{QCD}^{(4)}=215$~MeV
(326~MeV) for $n_f=4$ quark flavors, to be used in the one-loop (two-loop)
formula for $\alpha_s^{(n_f)}(\mu_r)$, with renormalization scale $\mu_r$.

\begin{figure*}
\centering
\includegraphics[width=4.4cm]{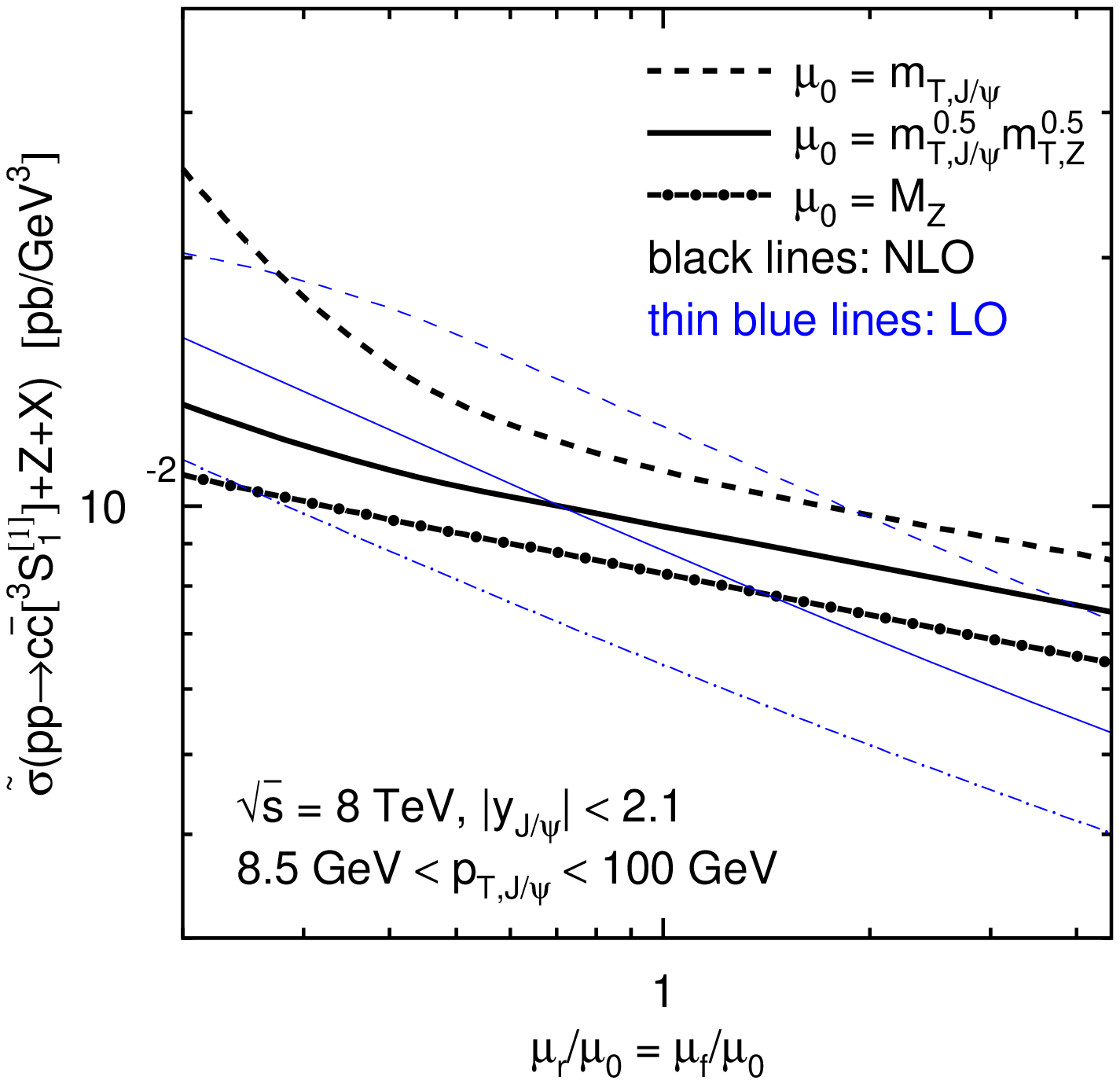}
\includegraphics[width=4.4cm]{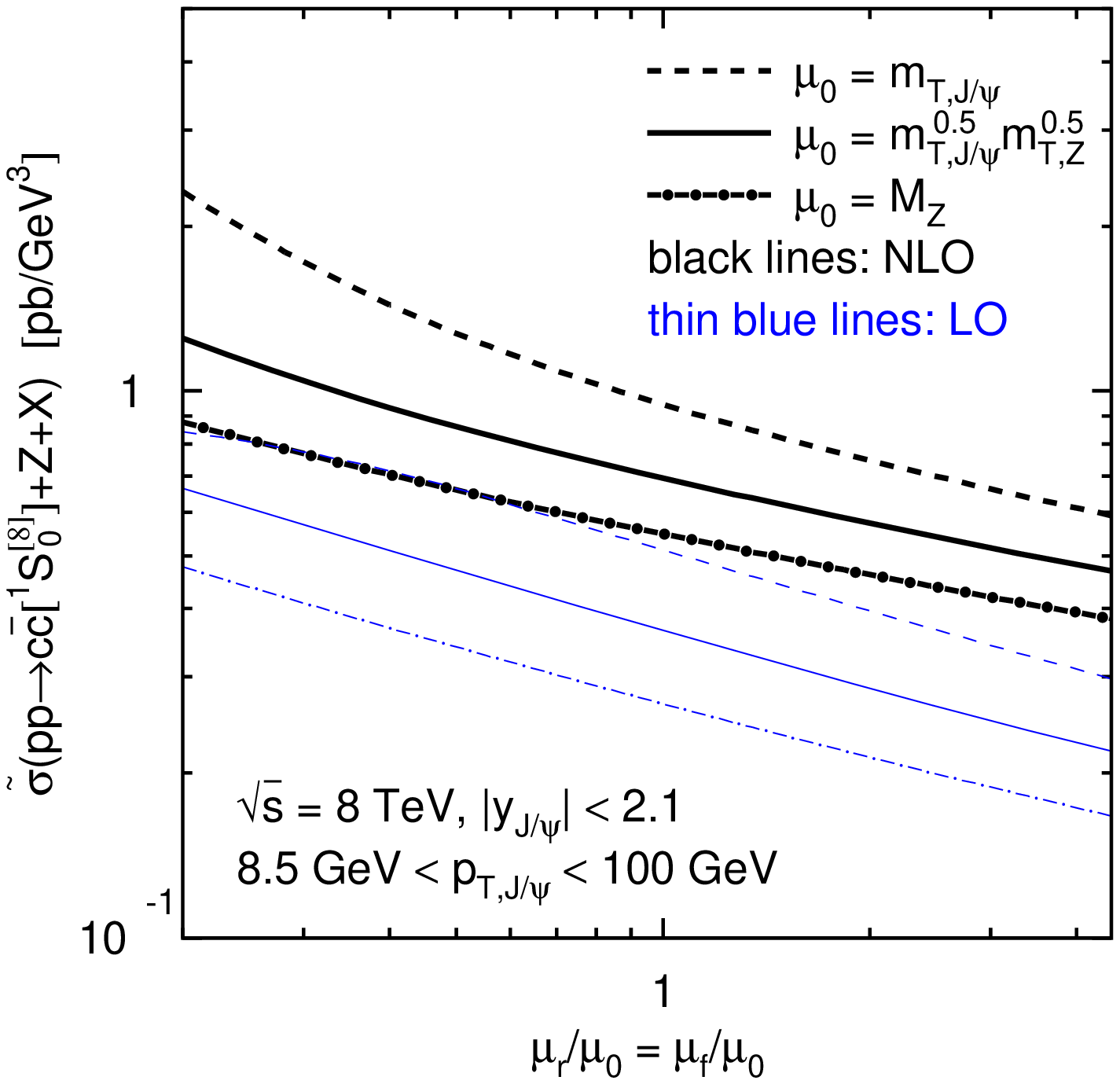}
\includegraphics[width=4.4cm]{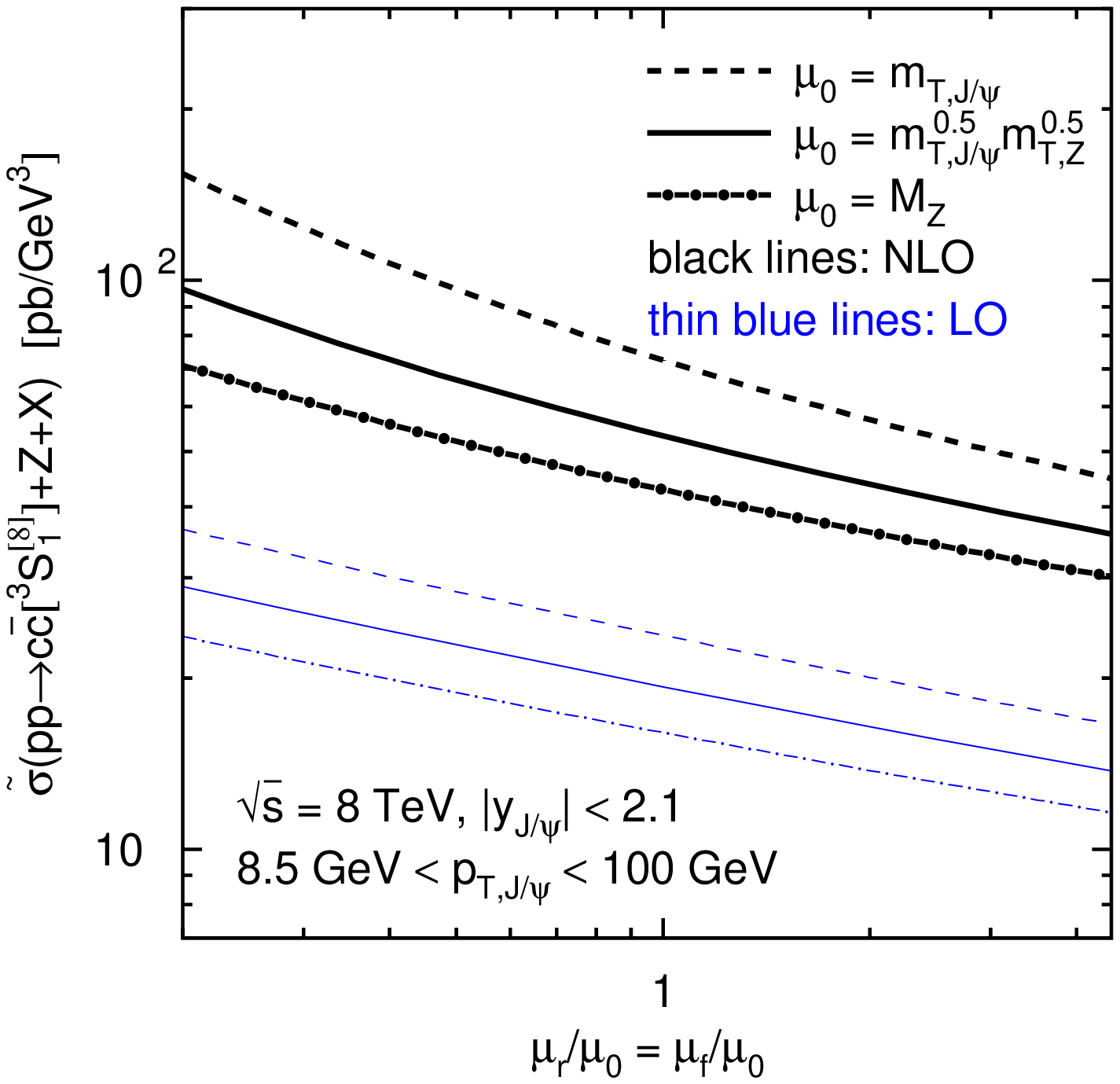}
\includegraphics[width=4.4cm]{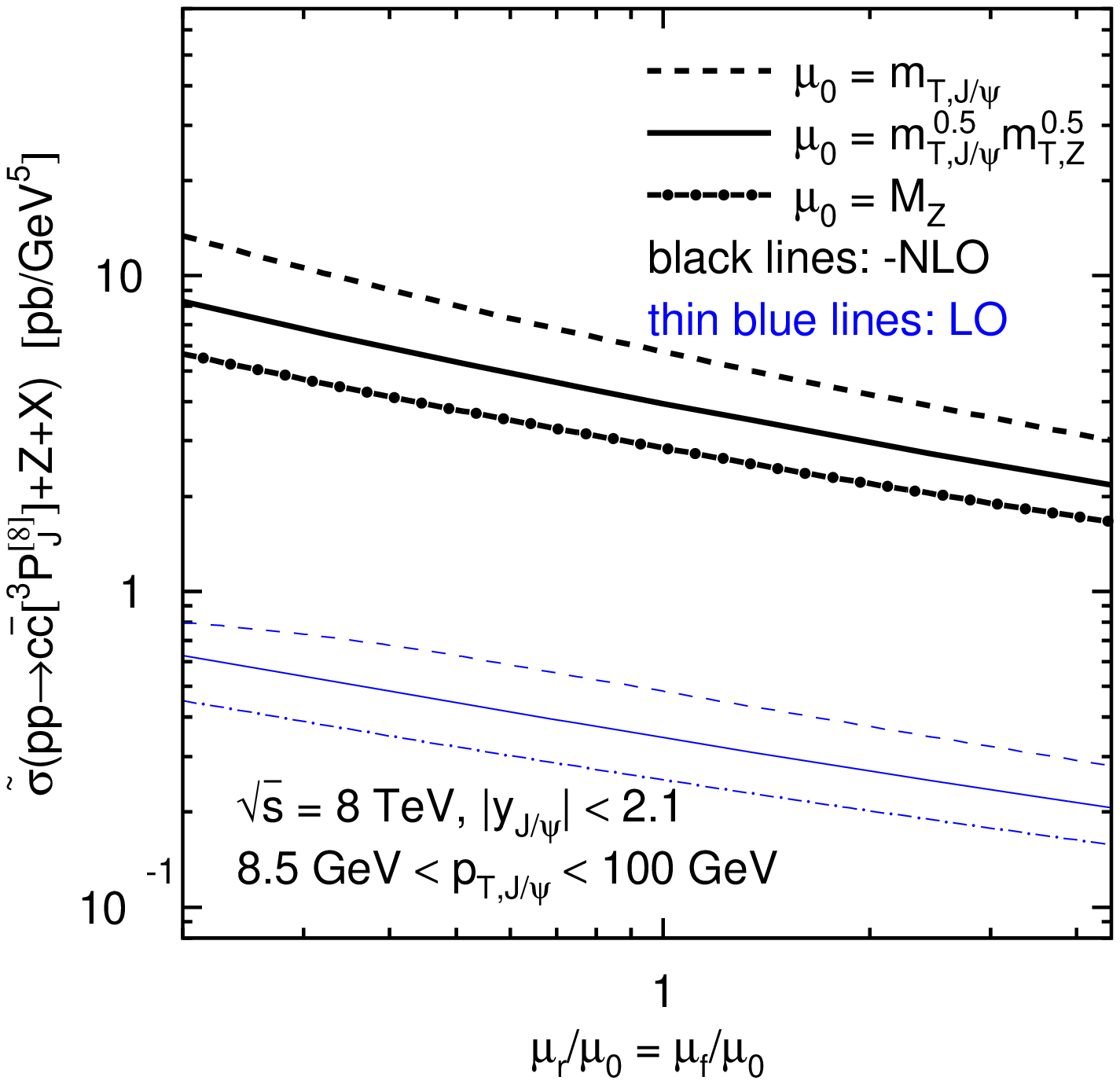}

\vspace{5pt}
\includegraphics[width=4.4cm]{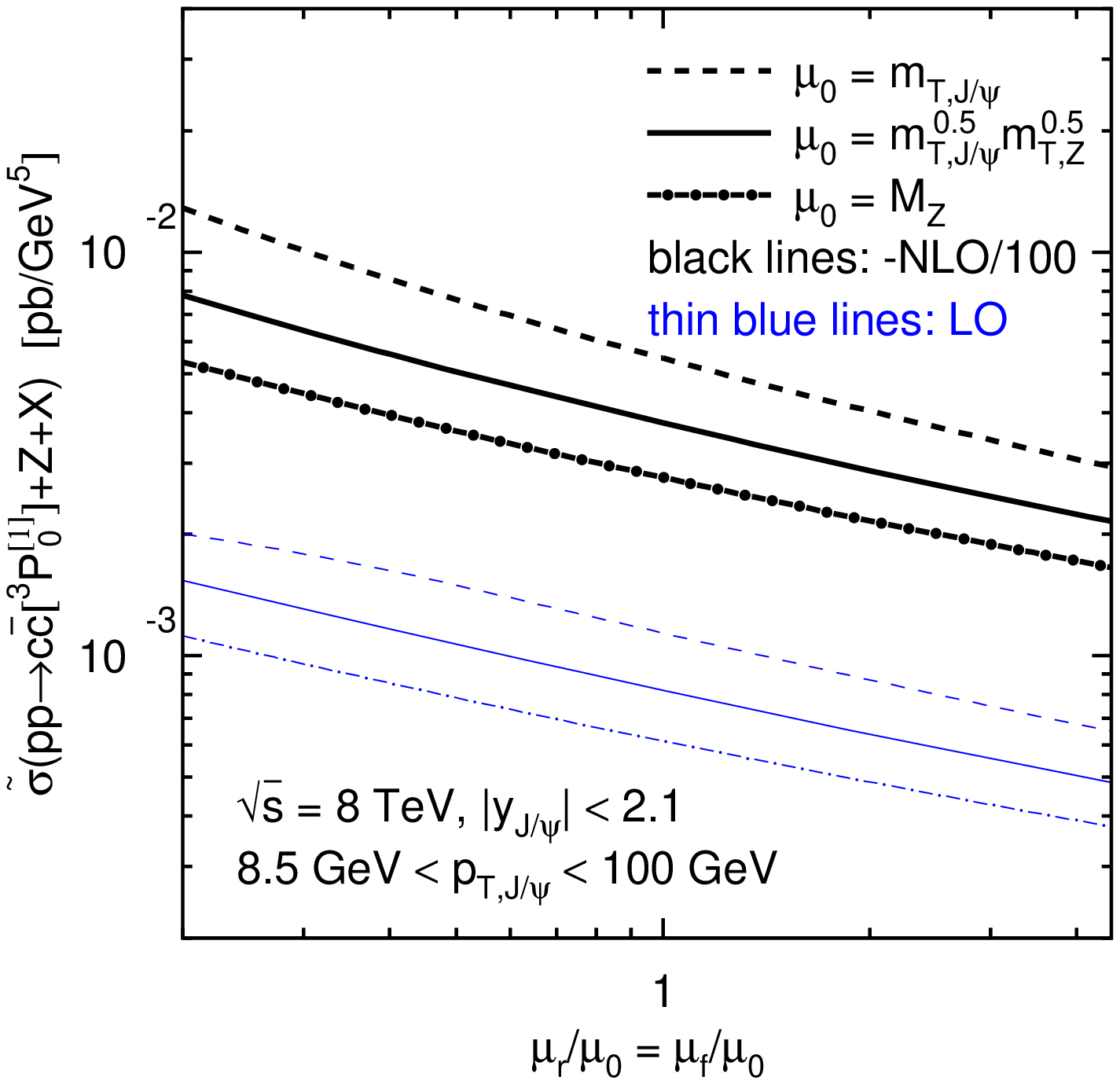}
\includegraphics[width=4.4cm]{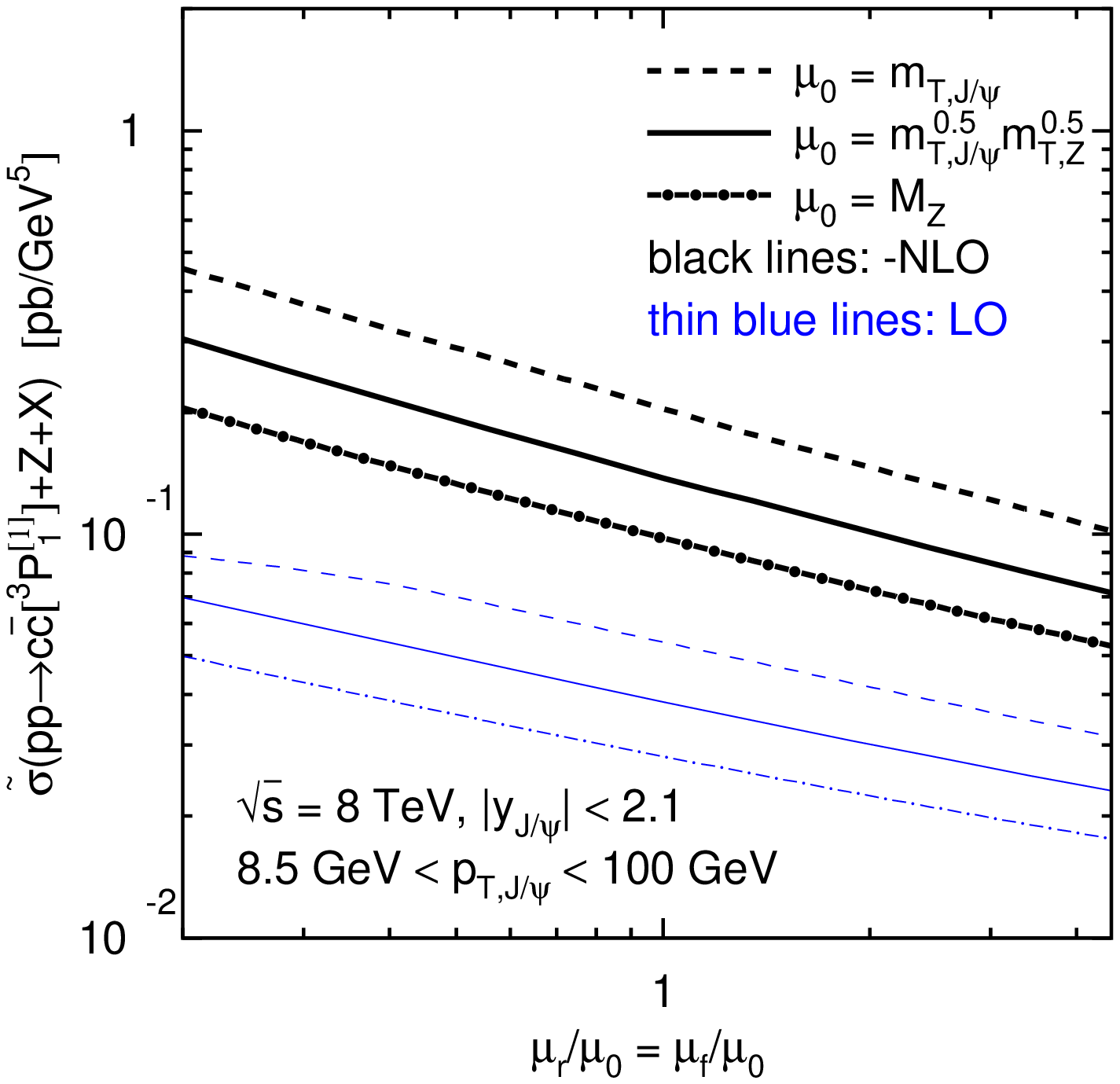}
\includegraphics[width=4.4cm]{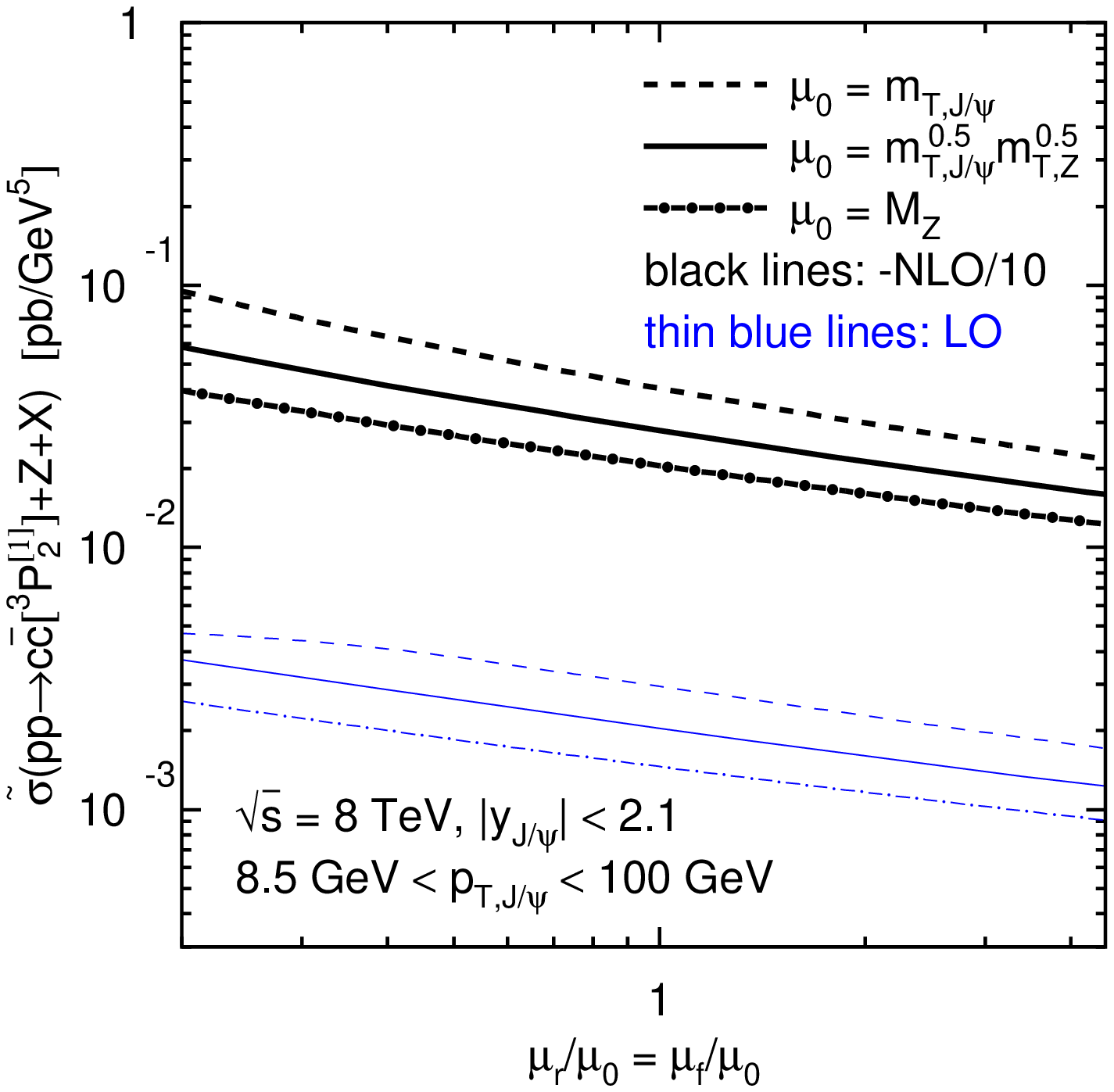}
\includegraphics[width=4.4cm]{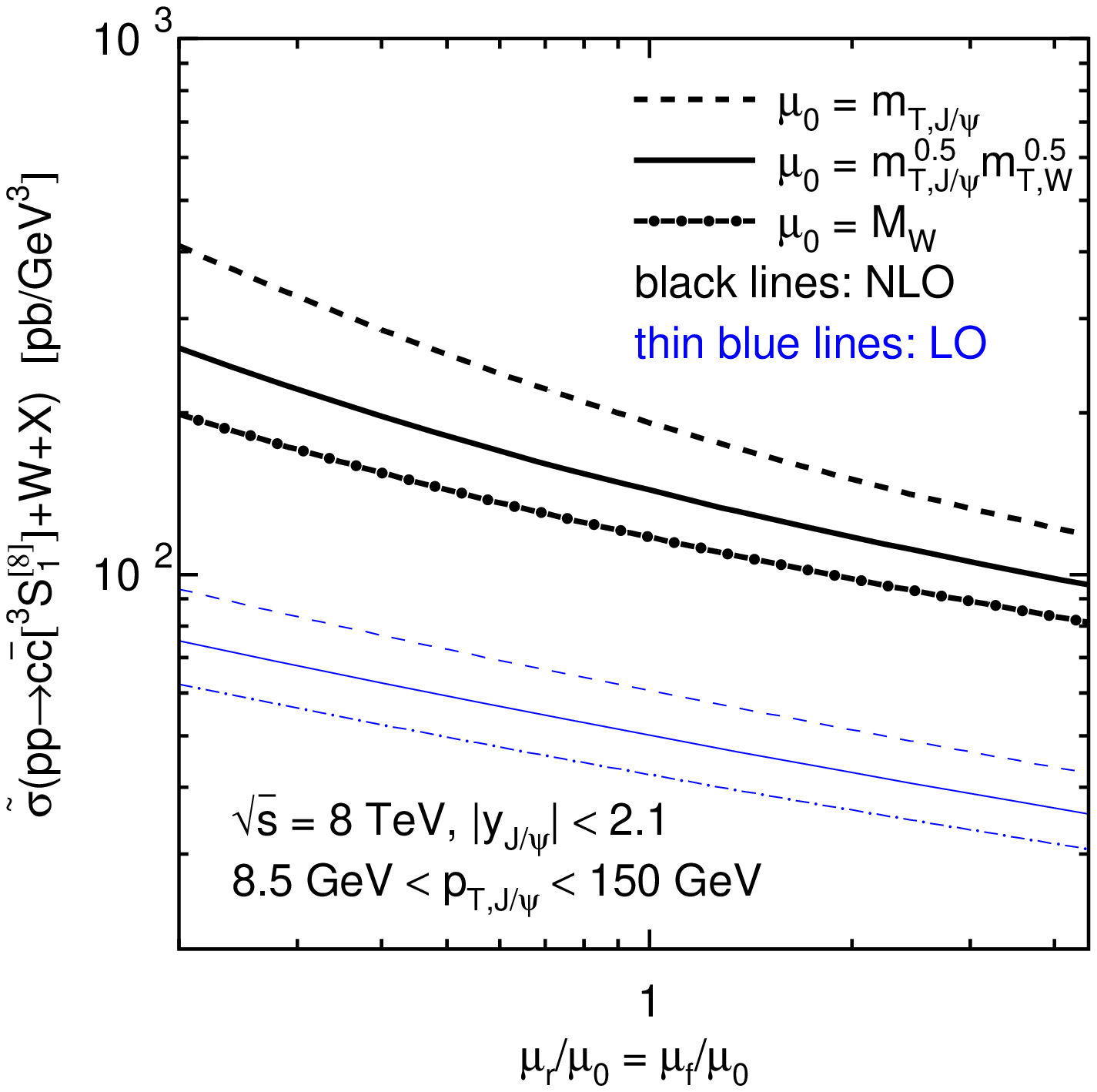}
\caption{\label{fig:scalevariation}%
  Dependences on $\mu/\mu_0$, for
  $\mu_0=m_{T,J/\psi},\sqrt{m_{T,J/\psi}m_{T,W/Z}},M_{W/Z}$,
  of the LO and NLO cross sections in Eq.~(\ref{eq:xs}) for ATLAS kinematics
  \cite{Aad:2014kba,Aaboud:2019wfr} selecting the Fock states $n$ that already
  contribute at LO.} 
\end{figure*}

Besides $\mu_r$, two more unphysical scales appear, namely, the
factorization scales of QCD and NRQCD, $\mu_f$ and $\mu_\Lambda$.
For definiteness, we put $\mu_\Lambda=m_c$ as default value and unify
$\mu=\mu_r=\mu_f$, for which plausible default choices include
$\mu_0=m_{T,J/\psi}$ \cite{Li:2010hc,Mao:2011kf},
$\mu_0=\sqrt{m_{T,J/\psi}m_{T,W/Z}}$ \cite{Kniehl:2002wd}, and
$\mu_0=M_{W/Z}$ \cite{Gong:2012ah},
where $m_T=\sqrt{m^2+p_T^2}$ is the transverse mass of a particle with mass $m$
and transverse momentum $p_T$.
In Fig.~\ref{fig:scalevariation}, we investigate, for each choice of $\mu_0$,
the dependencies on $\mu/\mu_0$ of the LO and NLO cross sections in
Eq.~(\ref{eq:xs}) for ATLAS kinematics \cite{Aad:2014kba,Aaboud:2019wfr}
selecting the Fock states $n$ that already contribute at LO.
We observe that the reduction in $\mu$ dependence when going from LO to NLO
is least favorable for $\mu_0=m_{T,J/\psi}$, which ignores the influence of
$M_W$ and $M_Z$ on the scale setting.
On the other hand, in the case of the important color singlet channel
$pp\to c\overline{c}[^3S_1^{[1]}]+Z+X$, the difference between the LO and NLO
cross sections is particularly small for the democratic choice
$\mu_0=\sqrt{m_{T,J/\psi}m_{T,W/Z}}$, which we thus adopt henceforth.

The experimental data as presented in
Refs.~\cite{Aad:2014kba,Aad:2014rua,Aaboud:2019wfr} are not directly suitable
for comparisons with our theoretical predictions.
Firstly, they include contributions from double parton scattering (DPS),
where two partons out of the same proton participate in the hard collision,
while our predictions only include single parton scattering (SPS).
Fortunately, in Refs.~\cite{Aad:2014kba,Aad:2014rua,Aaboud:2019wfr}, the DPS
contributions have been estimated for each bin, using as input the universal
DPS effective area $\sigma_\mathrm{eff}=15^{+5.8}_{-4.2}$~mb measured
in Ref.~\cite{Aad:2013bjm}, so that we can conveniently subtract them out from
the measured cross sections.
Secondly, the $J/\psi+W/Z+X$ cross section data are in Refs.~\cite{Aad:2014kba,Aad:2014rua,Aaboud:2019wfr} normalized  to the total cross sections, $\sigma_W=\sigma(pp\to W+X)$ and $\sigma_Z=\sigma(pp\to Z+X)$, respectively. To undo the normalization, in the $W$ case, we rely on the ATLAS \cite{Aaboud:2016btc} and CMS
\cite{Chatrchyan:2014mua} measurements of
$\sigma_W\times\mathrm{Br}(W\to l \nu)$ at $\sqrt{s}=7$~TeV and 8~TeV,
respectively.
In the $Z$ case, we resort to the CMS measurement of
$\sigma_Z\times\mathrm{Br}(Z\to l^+ l^-)$ at $\sqrt{s}=8$~TeV
\cite{Chatrchyan:2014mua}, which is, however, bound to include a non-negligible
number of $\gamma^*$ background events due to the relatively large
$l^+l^-$ invariant mass acceptance cut of 60~$\mathrm{GeV}<m_{l^+l^-}<120$~GeV.
This background has been estimated to be 3\% using Monte Carlo simulations
in Ref.~\cite{Chatrchyan:2014mua}.
On the other hand, thanks to the much tighter $m_{l^+l^-}$ cut, of just
$\pm10$~GeV around the $Z$ peak, the measurement of Ref.~\cite{Aad:2014kba}
should hardly be contaminated by $\gamma^*$ events.
To correct for this mismatch, we subtract 3\% from the result for
$\sigma_Z\times\mathrm{Br}(Z\to l^+ l^-)$ in Ref.~\cite{Chatrchyan:2014mua}.
To summarize, we have $\sigma_W=(98.71\pm2.34)$~nb at $\sqrt{s}=7$~TeV, and
$\sigma_W=(112.43\pm3.81)$~nb and $\sigma_Z=(33.14\pm1.19)$~nb at
$\sqrt{s}=8$~TeV.

We employ four popular NLO LDME sets, in which the CS LDMEs have been evaluated
using potential models or extracted from measured leptonic decay rates and the
CO LDMEs have been fitted to experimental data of single inclusive production,
with different data selections and fit strategies.
Set~1 is a combination of (i) the $J/\psi$ LDMEs obtained by a global
fit to prompt production data, with $p_T>1$~GeV for photoproduction and
two-photon scattering and $p_T>3$~GeV for hadroproduction, after subtracting
the estimated feed-down contributions \cite{Butenschoen:2011yh};
(ii) the $\psi(2S)$ LDMEs recently determined from a global fit to data of
unpolarized hadroproduction with $p_T>1$~GeV \cite{Butenschoen:2022orc};
and (iii) the $\chi_{cJ}$ LDMEs determined in Ref.~\cite{Ma:2010vd} from a fit
to Tevatron data with $p_T>4$~GeV of the $\chi_{c2}$ to $\chi_{c1}$ cross section
ratio.
Set~2 \cite{Gong:2012ug} has been fitted to prompt production data with
$p_T>7$~GeV from the Tevatron and the LHC.
Set~3 \cite{Bodwin:2015iua} has been fitted to prompt production data, with
$p_T>10$~GeV for $J/\psi$ mesons and $p_T>11$~GeV for $\chi_{cJ}$ and $\psi(2S)$
mesons, from the Tevatron and the LHC, combining fixed-order results with
fragmentation contributions computed in the leading-power factorization
formalism and thus resumming logarithms of $p_T^2/(2m_c)^2$.
Set~4 \cite{Brambilla:2022rjd} has been determined by a joint fit to LHC data
of prompt $J/\psi$, $\psi(2S)$, $\Upsilon(2S)$, and $\Upsilon(3S)$ production,
imposing $p_T>9$~GeV for charmonium and $p_T>28.5$~GeV for bottomonium,
subtracting estimated $\chi_{cJ}$ feed-down contributions, and implementing
constraints from a potential NRQCD analysis of the LDMEs.
Since Ref.~\cite{Brambilla:2022rjd} does not provide $\chi_{cJ}$ LDMEs, we set
them to zero keeping the omission of the $\chi_{cJ}$ feed-down contributions in
mind as an unaccounted source of systematic uncertainty.

\begin{table*}
\begin{center}
\begin{tabular}{l||ccccc|cccccc}
\multicolumn{1}{c}{} & \multicolumn{5}{c}{$pp\to c\overline{c}[n]+Z+X$} & \multicolumn{6}{c}{$pp\to c\overline{c}[n]+W+X$}
\\
$p_{T,J/\psi}$ $[\mathrm{GeV}]$ & 8.5 -- 10 & 10 -- 14 & 14 -- 18 & 18 -- 30 & 30 -- 100 & 8.5 -- 10 & 10 -- 14 & 14 -- 18 & 18 -- 30 & 30 -- 60 & 60 -- 150 \\
\hline
\hline
$n={^3}S_1^{[1]}$, LO &
$0.0862$ &
$0.0488$ &
$0.0228$ &
$0.00731$ &
$0.000334$ &
$0$ &
$0$ &
$0$ &
$0$ &
$0$ &
$0$\\
$n={^3}S_1^{[1]}$, NLO &
$0.0806$ &
$0.0489$ &
$0.0251$ &
$0.00906$ &
$0.000558$ &
$0$ &
$0$ &
$0$ &
$0$ &
$0$ &
$0$\\
\hline
$n={^1}S_0^{[8]}$, LO  &
$3.07$ &
$1.91$ &
$0.985$ &
$0.349$ &
$0.0202$ &
$0$ &
$0$ &
$0$ &
$0$ &
$0$ &
$0$\\
$n={^1}S_0^{[8]}$, NLO &
$5.88$ &
$3.56$ &
$1.80$ &
$0.648$ &
$0.0443$ &
$8.49$ &    
$4.59$ &    
$2.11$ &    
$0.722$ &   
$0.111$ &   
$0.00604$ \\
\hline
$n={^3}S_1^{[8]}$, LO &
$127$ &
$83.7$ &
$48.6$ &
$21.7$ &
$2.48$&
$338$ &
$220$ &
$126$ &
$55.0$ &
$11.8$ &
$0.919$ \\
$n={^3}S_1^{[8]}$, NLO &
$365$ &
$236$ &
$134$ &
$58.6$ &
$6.62$&
$1000$ & 
$637$ &  
$361$ &  
$155$ &  
$33.2$ & 
$2.89$ \\
\hline
$n={^3}P_J^{[8]}$, LO &
$2.86$ &
$1.77$ &
$0.923$ &
$0.335$ &
$0.0208$ &
$0$ &
$0$ &
$0$ &
$0$ &
$0$ &
$0$ \\
$n={^3}P_J^{[8]}$, NLO &
$-23.8$ &
$-16.2$ &
$-9.72$ &
$-4.59$ &
$-0.564$&
$-72.0$ &  
$-47.9$ &  
$-28.3$ &  
$-12.6$ &  
$-2.81$ &  
$-0.221$ \\
\hline
$n={^3}P_0^{[1]}$, LO &
$0.00472$ &
$0.00339$ &
$0.00220$ &
$0.00107$ &
$0.0000942$ &
$0$ &
$0$ &
$0$ &
$0$ &
$0$ &
$0$\\
$n={^3}P_0^{[1]}$, NLO &
$-2.45$ &
$-1.63$ &
$-0.948$ &
$-0.423$ &
$-0.0477$ &
$-6.46$ &   
$-4.31$ &   
$-2.49$ &   
$-1.09$ &   
$-0.230$ &  
$-0.0172$ \\
\hline
$n={^3}P_1^{[1]}$, LO &
$0.323$ &
$0.206$ &
$0.106$ &
$0.0363$ &
$0.00174$ &
$0$ &
$0$ &
$0$ &
$0$ &
$0$ &
$0$\\
$n={^3}P_1^{[1]}$, NLO &
$-0.751$ &
$-0.526$ &
$-0.340$ &
$-0.171$ &
$-0.0239$ &
$-2.46  $ &
$-1.75  $ &
$-1.09  $ &
$-0.532 $ &
$-0.127 $ &
$-0.0105$\\
\hline
$n={^3}P_2^{[1]}$, LO &
$0.0290$ &   
$0.0117$ &   
$0.00387$ &  
$0.00102$ &  
$0.0000531$ &
$0$ &
$0$ &
$0$ &
$0$ &
$0$ &
$0$\\
$n={^3}P_2^{[1]}$, NLO &
$-1.79$ &
$-1.19$ &
$-0.699$ &
$-0.317$ &
$-0.0371$ &
$-4.87 $ &
$-3.22 $ &
$-1.86 $ &
$-0.823$ &
$-0.178$ &
$-0.0138$\\
\end{tabular}
\end{center}
\caption{\label{tab:sdcs}%
LO and NLO cross sections
$d\tilde{\sigma}(pp\to c\overline{c}[n]+W/Z+X)/dp_{T,J/\psi}
\times{\rm Br}(J/\psi\to\mu^+\mu^-)$
of Eq.~(\ref{eq:xs}) for all contributing Fock states $n$
in fb/GeV$^4$ (fb/GeV$^6$) for $S$ ($P$) wave states,
assuming the ATLAS kinematic conditions at $\sqrt{s}=8$~TeV
\cite{Aad:2014kba,Aaboud:2019wfr} including the binning in $p_{T,J/\psi}$.
The common shorthand notation
$d\tilde{\sigma}(pp\to c\overline{c}[{^3P}_J^{[8]}]+W/Z+X)$ implies
$\sum_{J=0}^2 (2J+1) d\tilde{\sigma}(pp\to c\overline{c}[{^3P}_{J}^{[8]}]+W/Z+X)$.
The integration accuracy is around 1\%.} 
\end{table*}

To enable interested readers to perform comparisons with alternative LDME
sets, we list in Table~\ref{tab:sdcs} the LO and NLO default cross sections
$d\tilde{\sigma}(pp\to c\overline{c}[n]+W/Z+X)/dp_{T,J/\psi}
\times{\rm Br}(J/\psi\to\mu^+\mu^-)$
of Eq.~(\ref{eq:xs}) assuming the ATLAS kinematic setup at $\sqrt{s}=8$~TeV
\cite{Aad:2014kba,Aaboud:2019wfr} including the binning in $p_{T,J/\psi}$.
Figure~\ref{fig:scalevariation} and Table~\ref{tab:sdcs} also usefully portray
the anatomy of the NLO corrections in the various $n$ channels as for sign and
magnitude.
The NLO NRQCD predictions are likely to be more reliable in the $Z$ case than
in the $W$ case, where we expect large next-to-next-to-leading-order
contributions due to the delayed unfolding of the $n$ structure, with only
$n={}^3S_1^{[8]}$ being present at LO and $n={}^3S_1^{[1]}$ not even at NLO.

\begin{figure*}
\centering
\includegraphics[width=4.4cm]{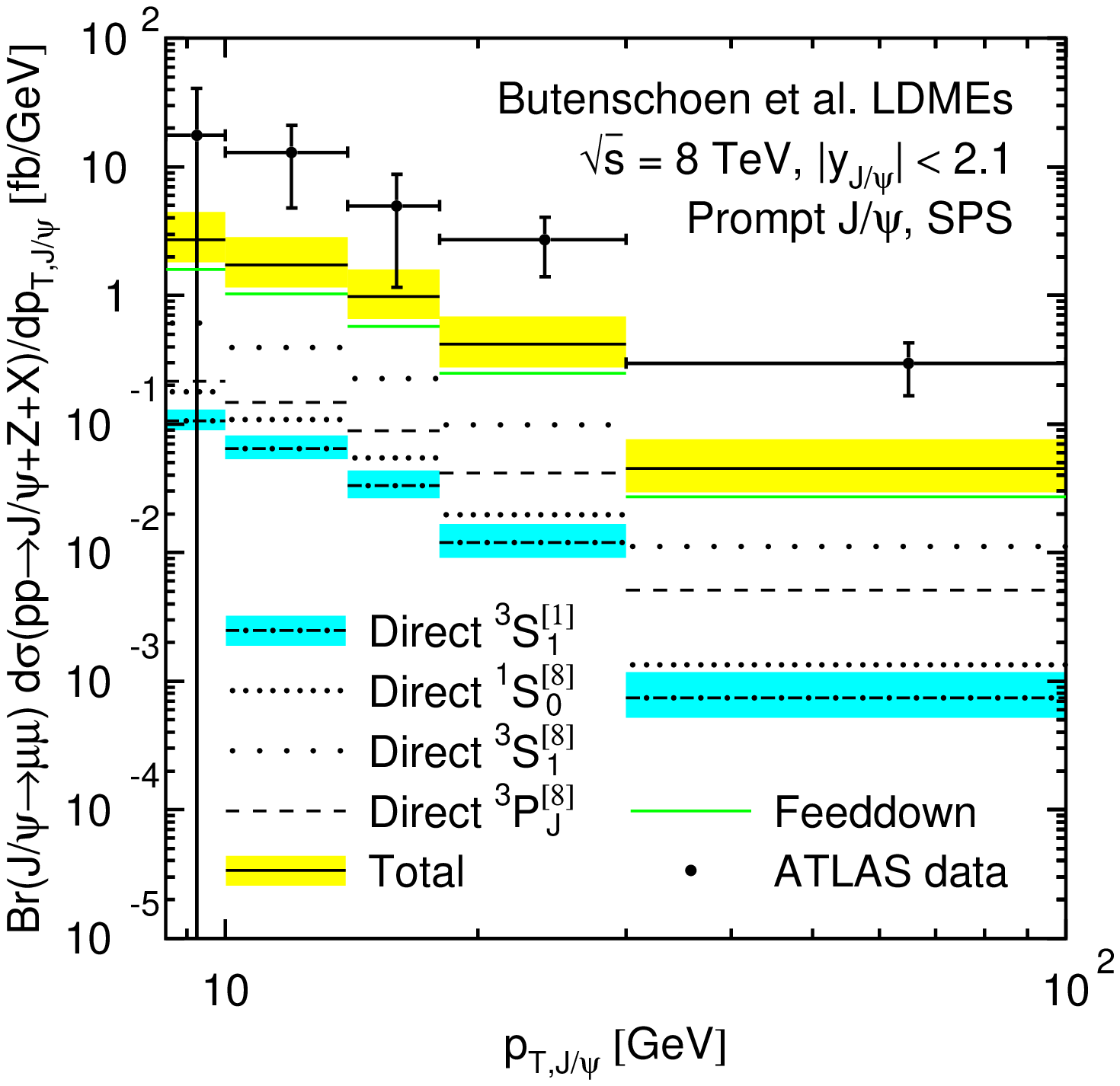}
\includegraphics[width=4.4cm]{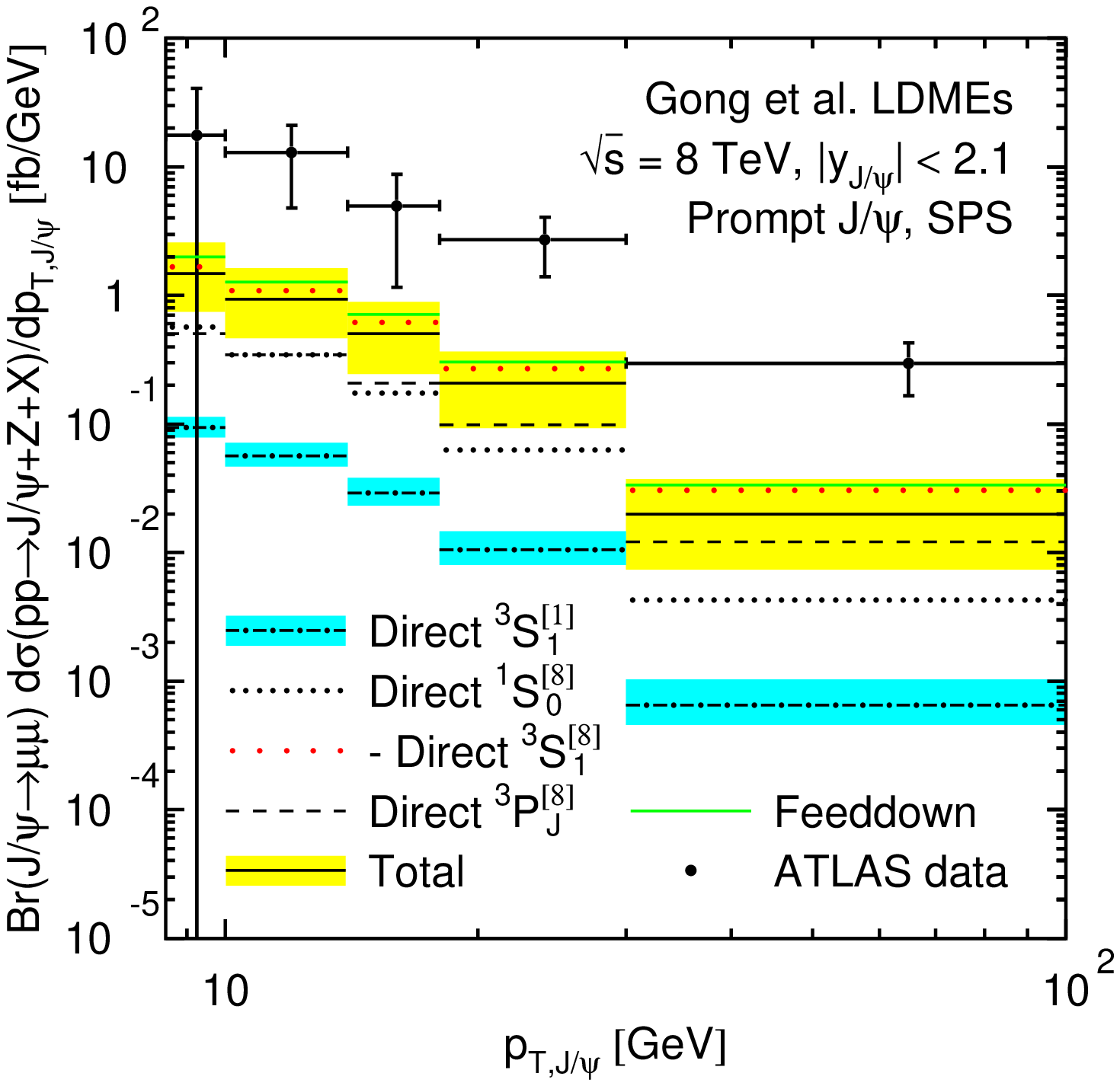}
\includegraphics[width=4.4cm]{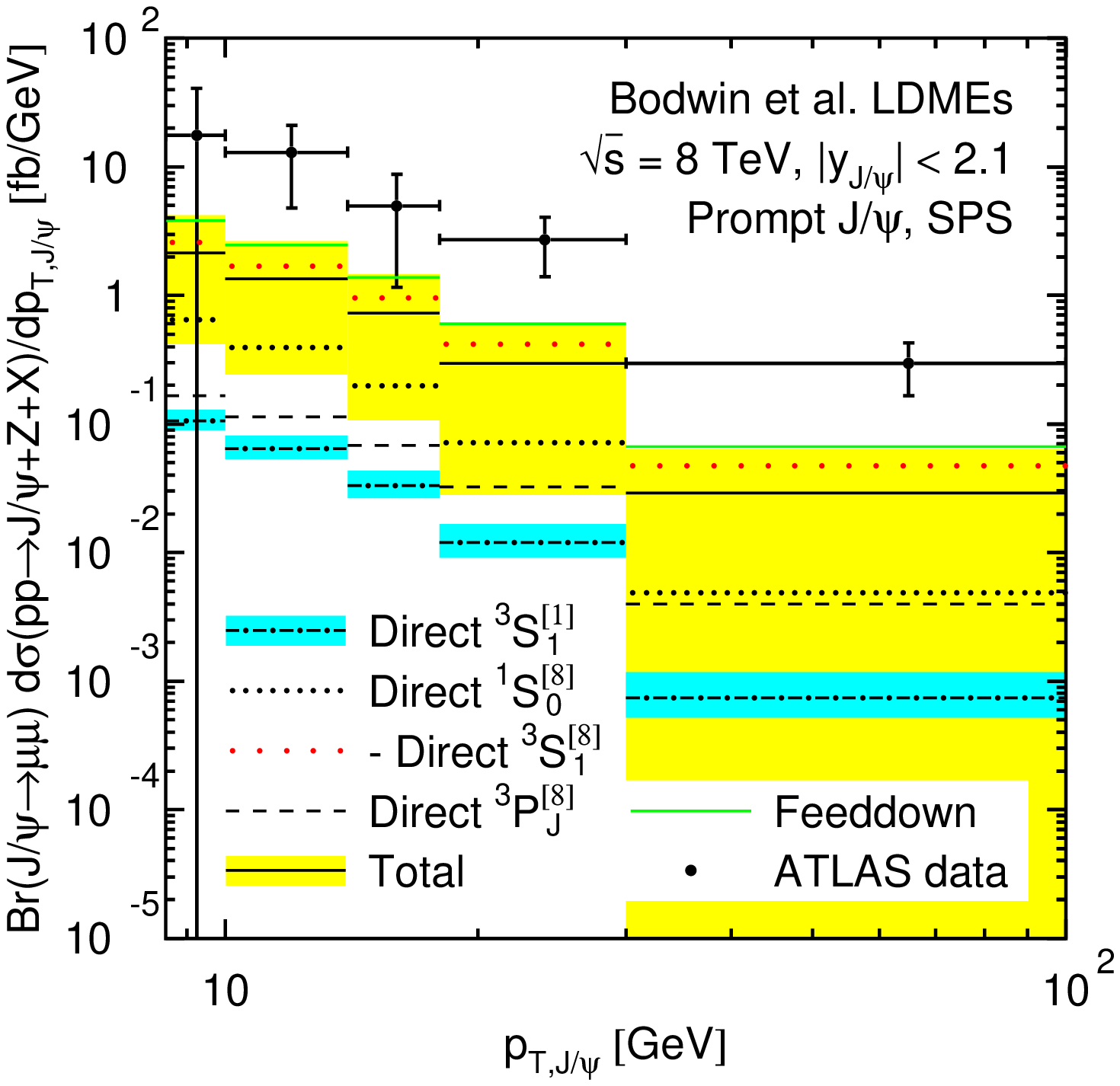}
\includegraphics[width=4.4cm]{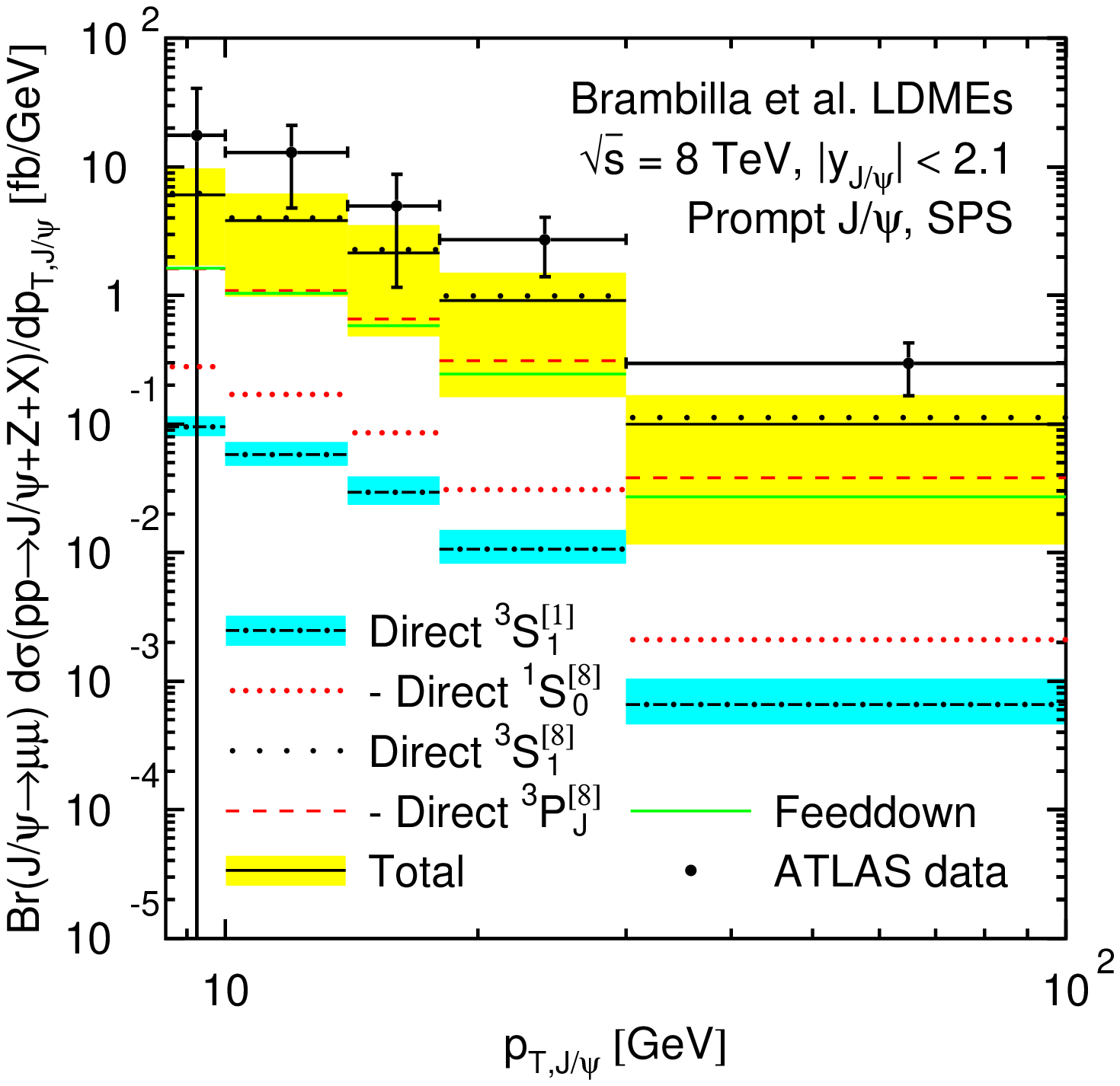}

\vspace{5pt}
\includegraphics[width=4.4cm]{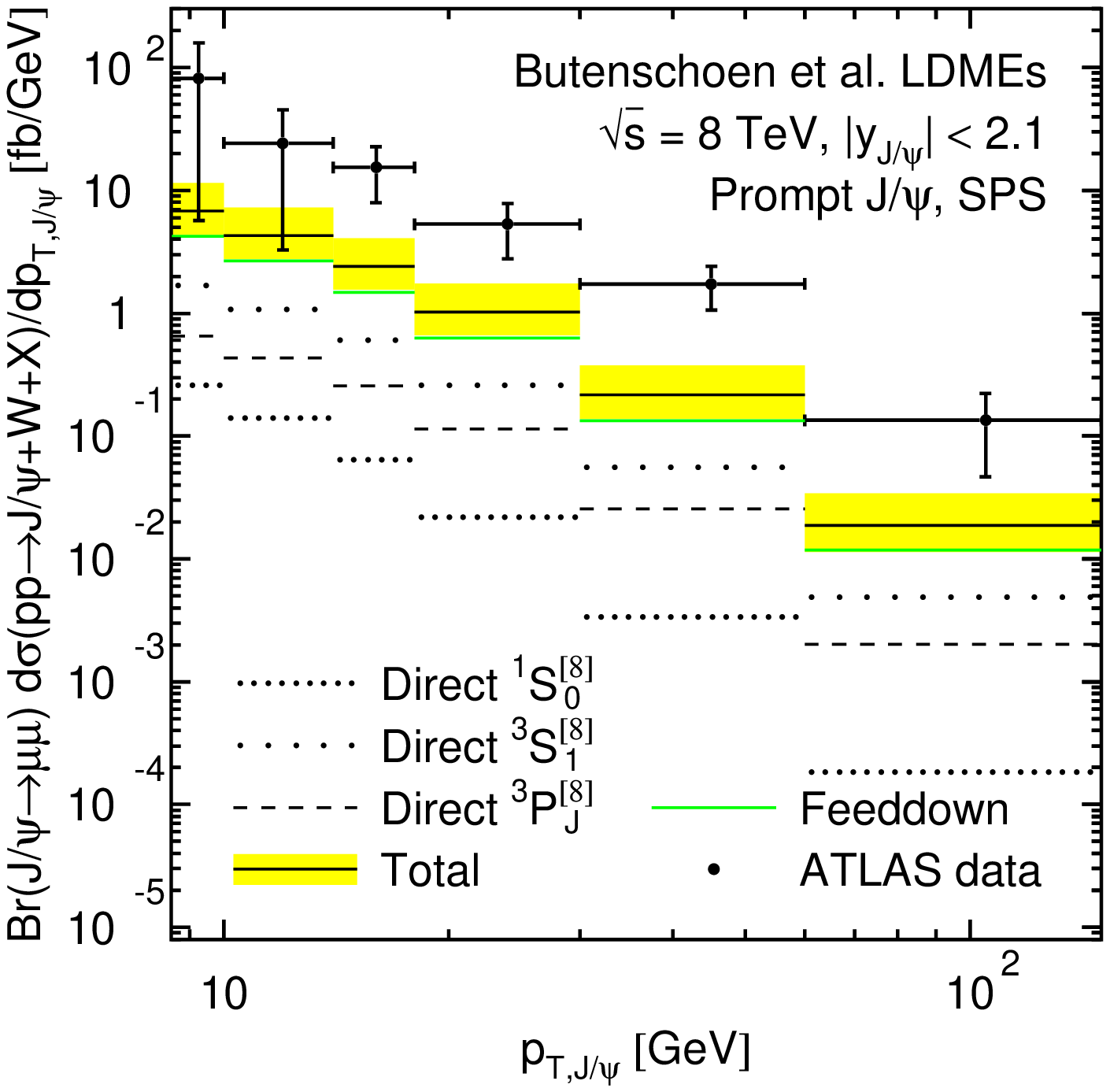}
\includegraphics[width=4.4cm]{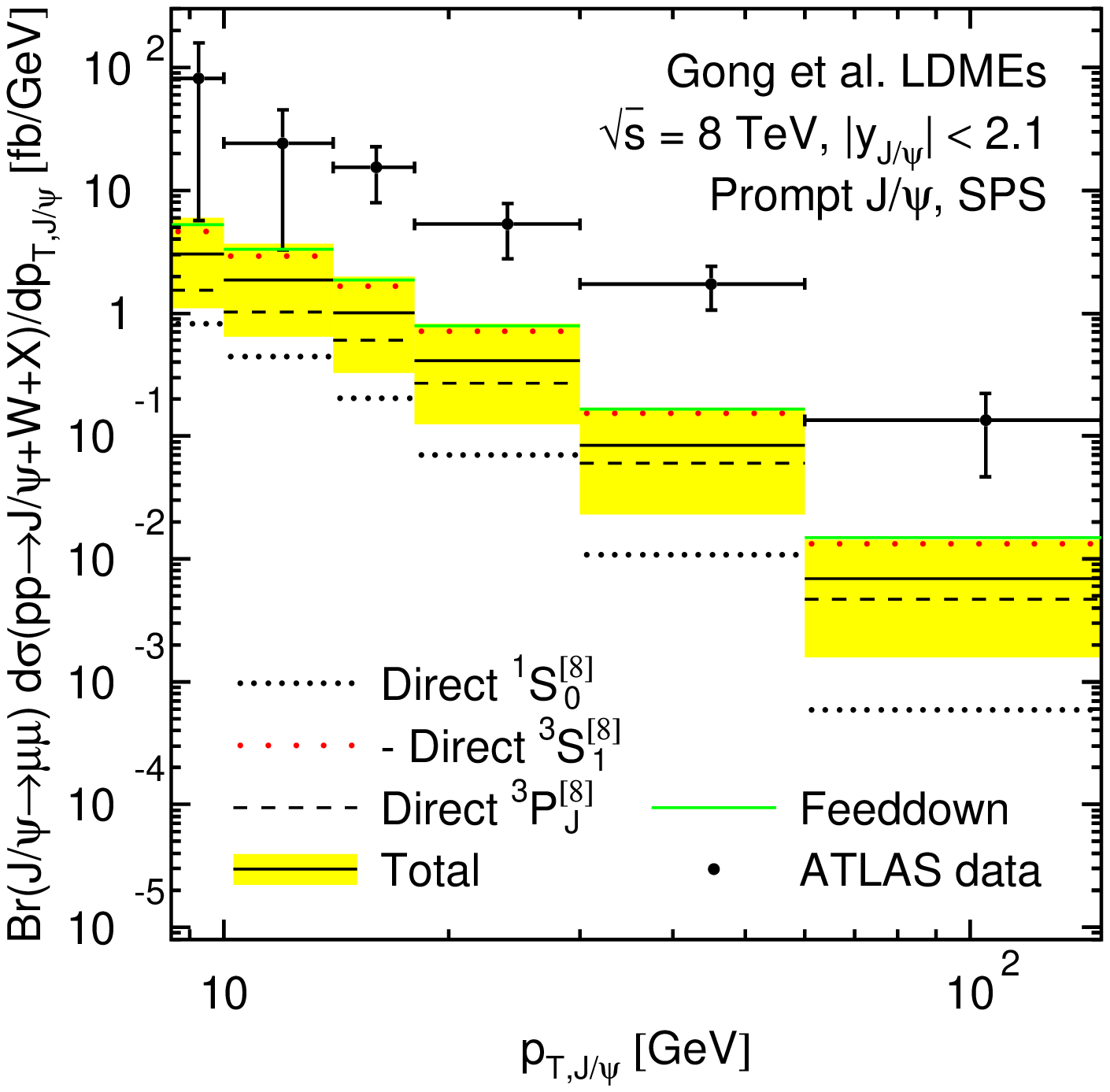}
\includegraphics[width=4.4cm]{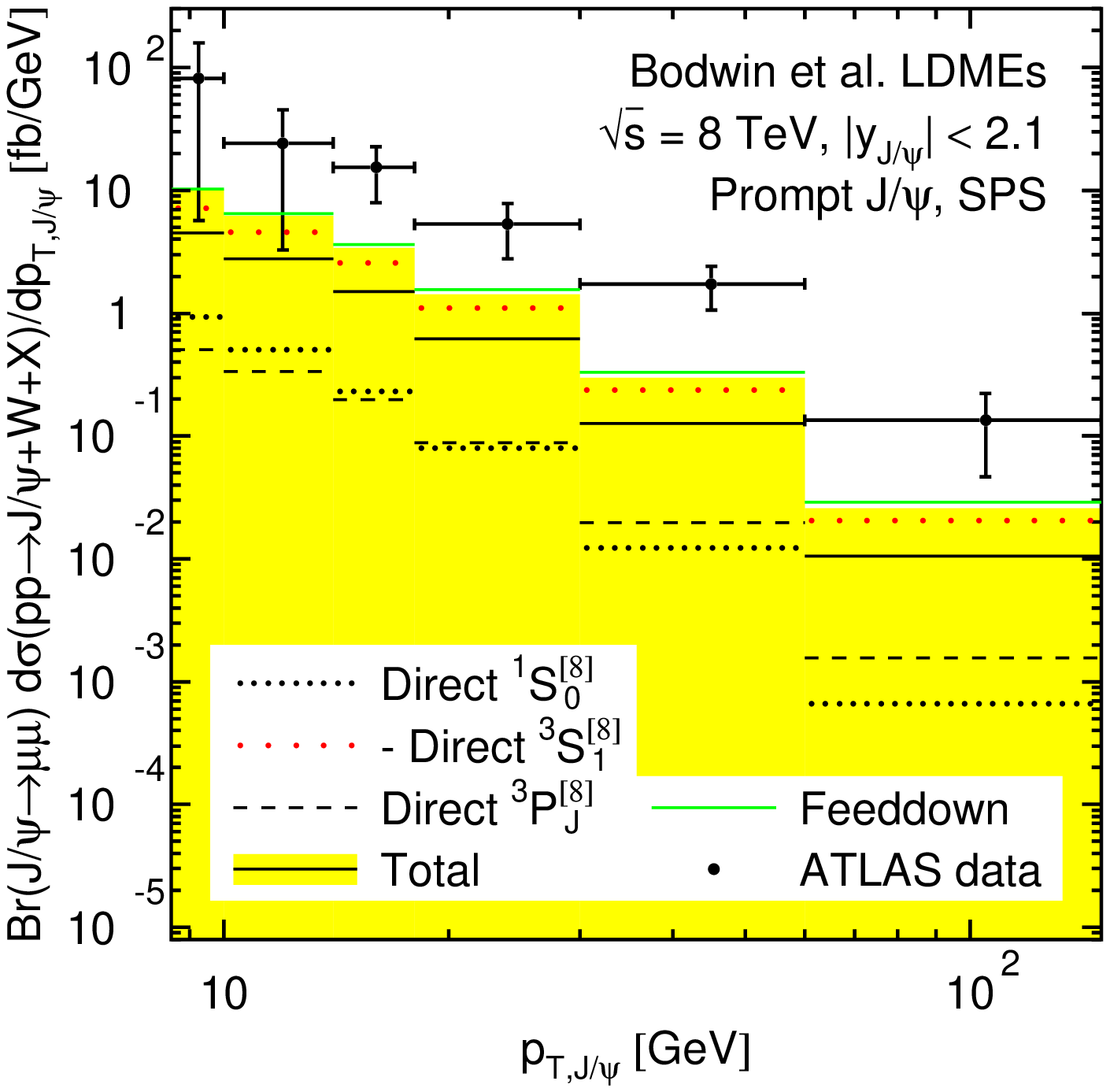}
\includegraphics[width=4.4cm]{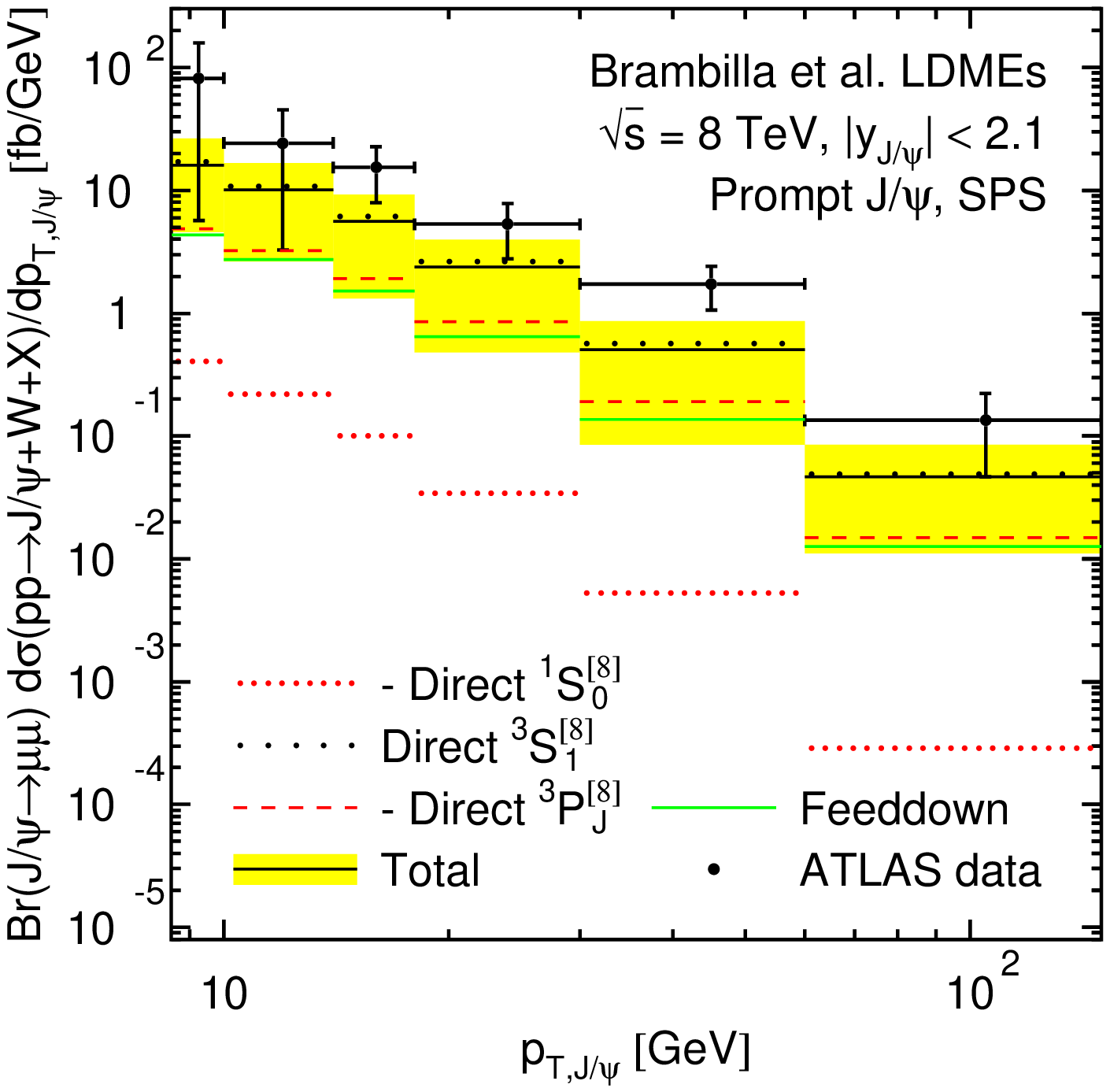}

\vspace{5pt}
\includegraphics[width=4.4cm]{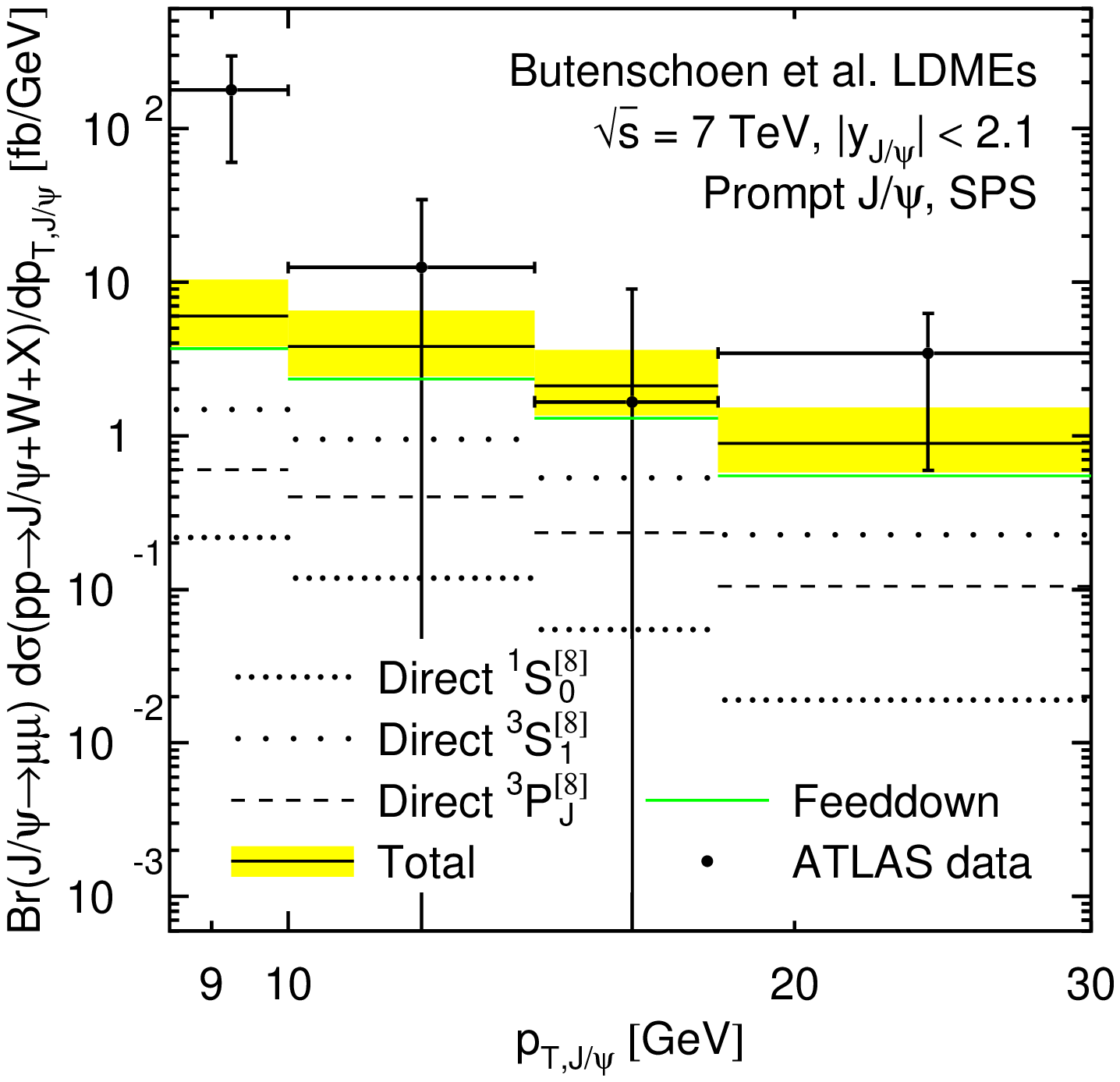}
\includegraphics[width=4.4cm]{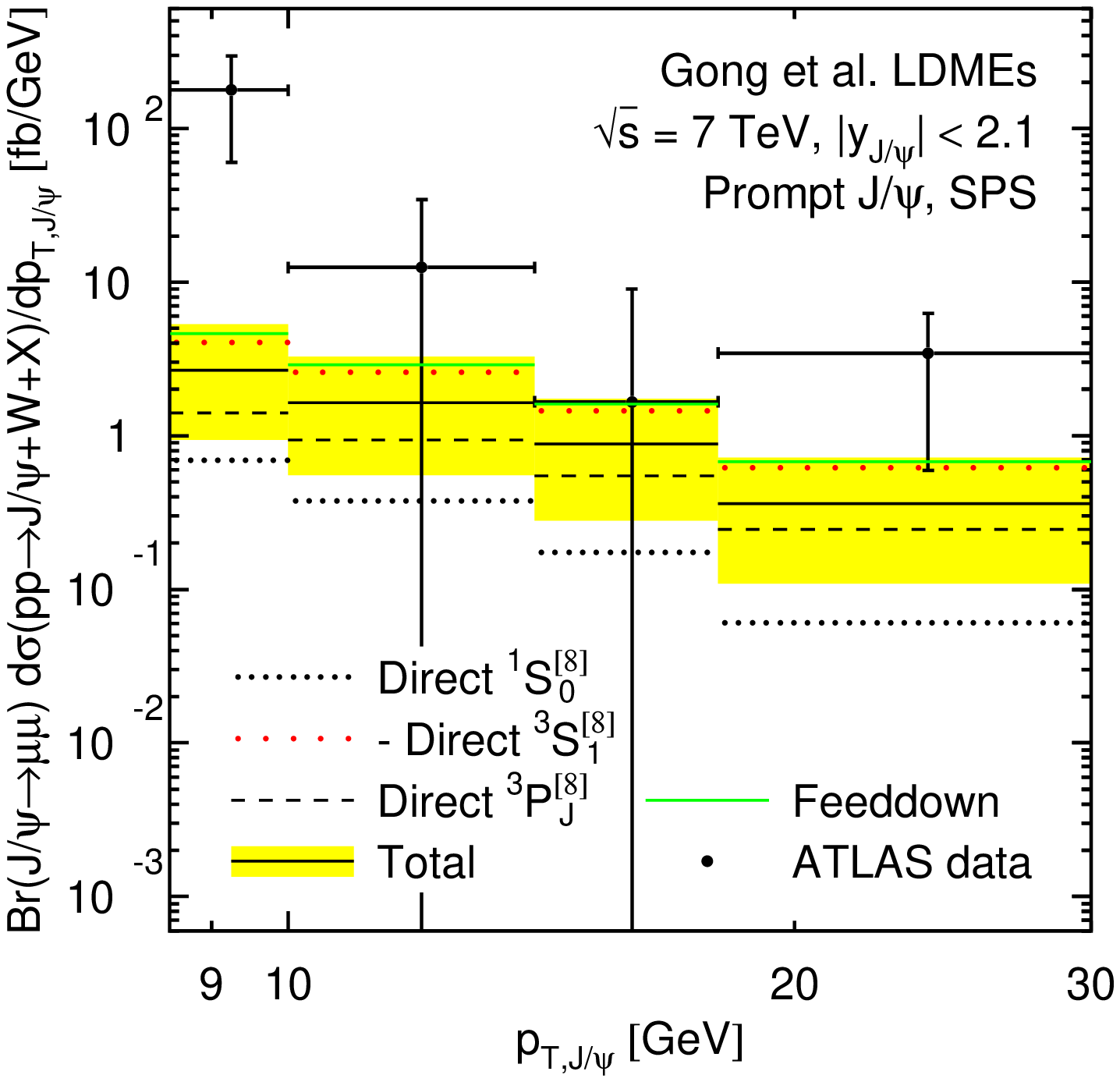}
\includegraphics[width=4.4cm]{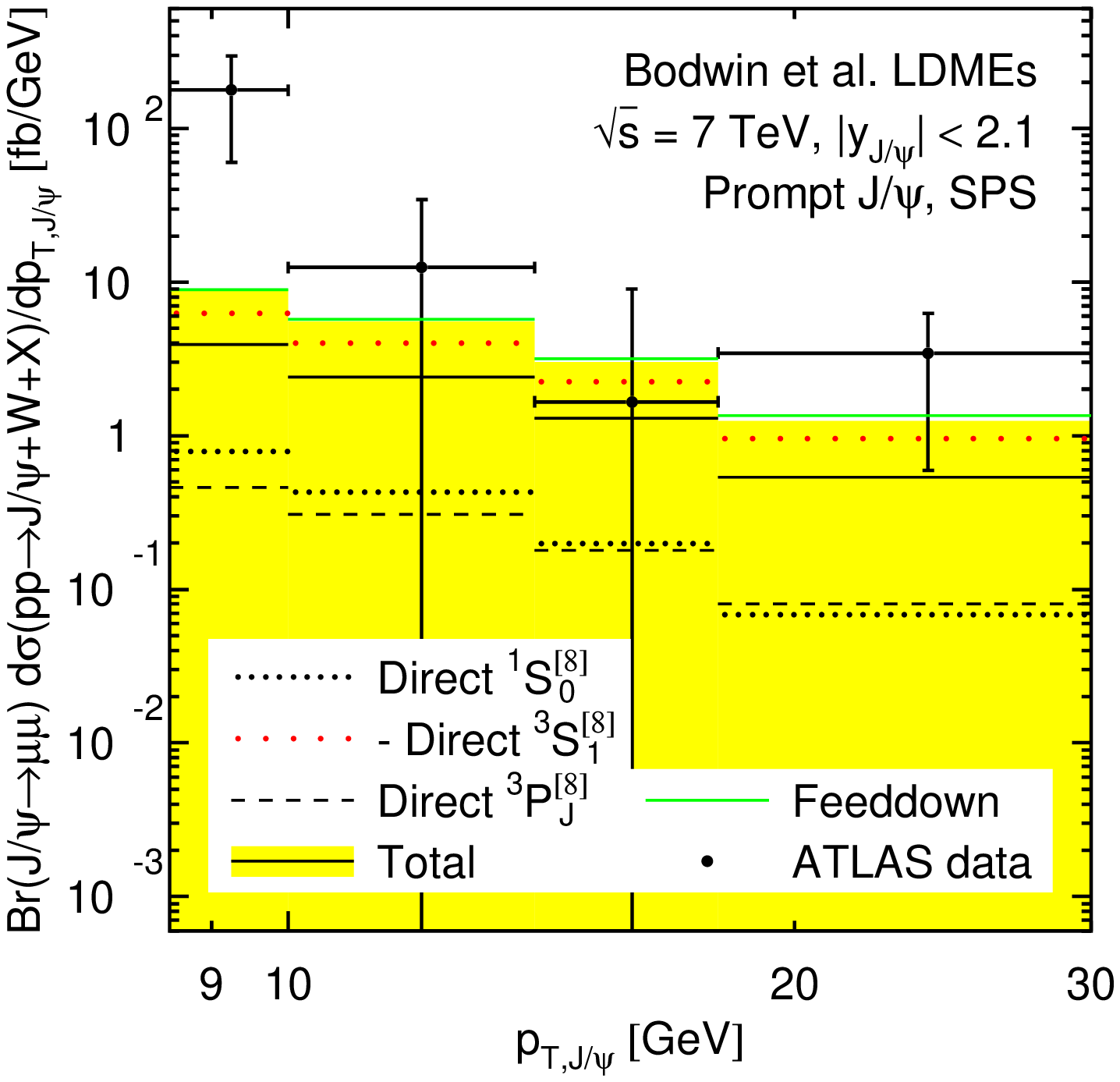}
\includegraphics[width=4.4cm]{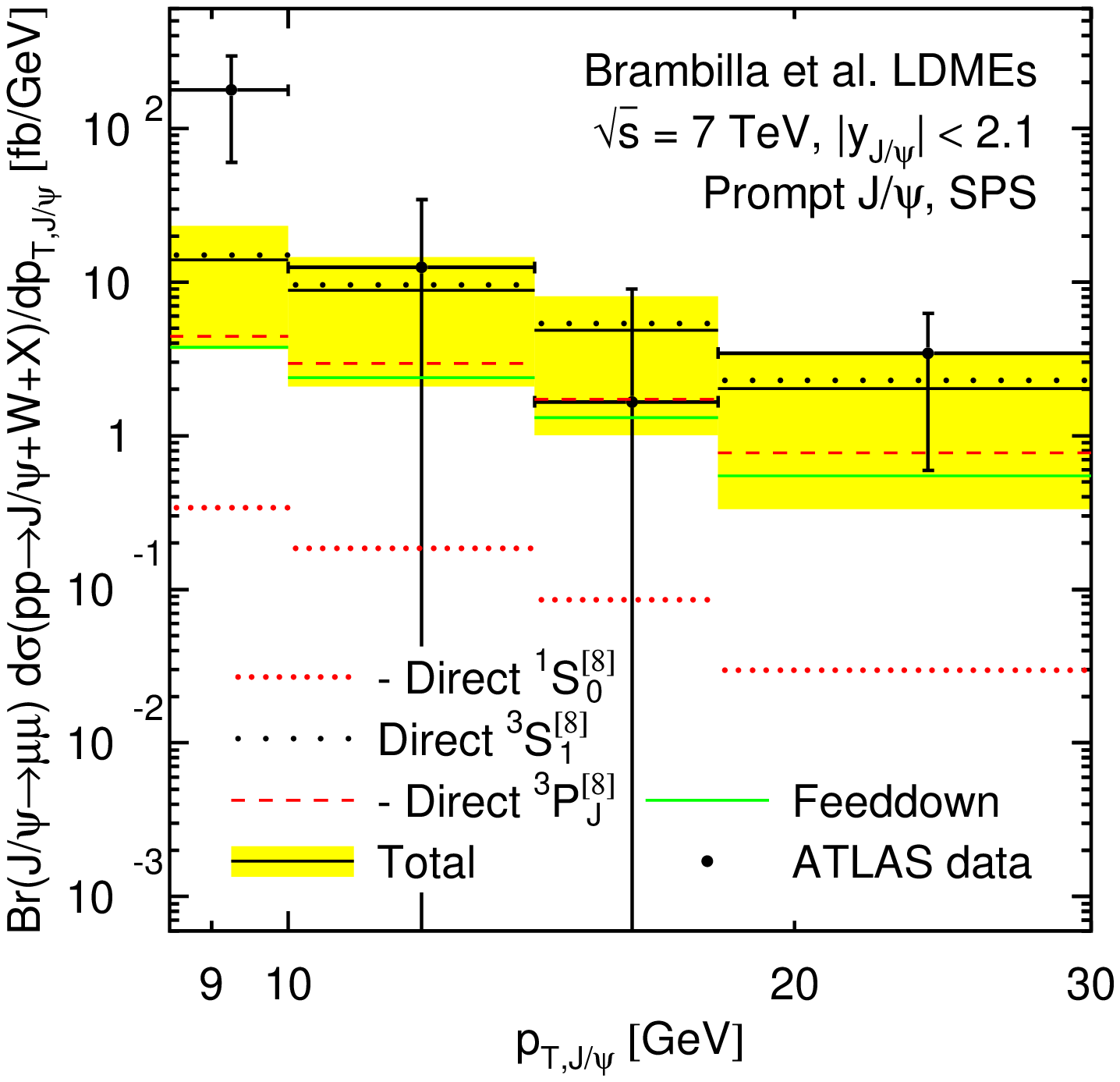}
\caption{\label{fig:results}%
  Comparison of the ATLAS data from
  Refs.~\cite{Aad:2014kba,Aaboud:2019wfr,Aad:2014rua} (rows), adjusted as
  described in the text, to our NLO predictions for
  $d\sigma(pp\to J/\psi+W/Z+X)/dp_{T,J/\psi}\times{\rm Br}(J/\psi\to\mu^+\mu^-)$
  in fb/GeV evaluated successively with LDME sets 1--4 (columns).
  The theoretical-uncertainty bands are evaluated as described in the text.}
\end{figure*}

In Fig.~\ref{fig:results}, we compare the ATLAS data
\cite{Aad:2014kba,Aaboud:2019wfr,Aad:2014rua}, modified as explained above, to
our NLO predictions for
$d\sigma(pp\to J/\psi+W/Z+X)/dp_{T,J/\psi}\times{\rm Br}(J/\psi\to\mu^+\mu^-)$
with the same binning in $p_{T,J/\psi}$.
The three rows in Fig.~\ref{fig:results} correspond to $J/\psi+Z$ production at
$\sqrt{s}=8$~TeV \cite{Aad:2014kba} and $J/\psi+W$ production at 8~TeV
\cite{Aaboud:2019wfr} and 7~TeV \cite{Aad:2014rua}, the four columns to LDME
sets 1--4.
In each frame, we break down the total result into the contributions from the
individual channels $n$ of direct production and the combined feed-down
contribution, and indicate theoretical uncertainties in the CSM and NRQCD
results.
The theoretical uncertainties are evaluated by adding in quadrature the errors
from the following three sources:
(i) variation of $\mu$ by a factor of 4 up and down relative to
$\mu_0=\sqrt{m_{T,J/\psi}m_{T,W/Z}}$;
(ii) variation of $\mu_\Lambda$ by a factor of 2 up and down relative to
$m_c$;
(iii) quadratic combination of the individual LDME errors quoted in
Refs.~\cite{Butenschoen:2011yh,Butenschoen:2022orc,Ma:2010vd,Gong:2012ug,Bodwin:2015iua,Brambilla:2022rjd}, making full use of the covariance matrices
available from
Refs.~\cite{Butenschoen:2011yh,Butenschoen:2022orc,Bodwin:2015iua,Brambilla:2022rjd}.
The large $\mu$ variation is to at least partially account for the
fact that also $\mu_0=M_{W/Z}$ is a plausible reference scale.

The default LO predictions, omitted in Fig.~\ref{fig:results} for clarity, may
be readily retrieved from Table~\ref{tab:sdcs}.
The $K$ factors at the bin level range between 0.9 and 1.7 in the CSM and
between 1.7 and 2.9 (1.8 and 4.8) in full NRQCD for the $Z$ ($W$) case,
underpinning the above expectation regarding the speed of convergence of the
perturbative expansions in both cases. Measuring the default NLO corrections
in terms of the LO standard deviations, we find the ranges $-0.10$ -- 0.91 in
the CSM, and 1.3 -- 3.2 (1.5 -- 5.0) in full NRQCD for the $Z$ ($W$) case.
The LO and NLO error bands always overlap at least partially, except for LDME
set~1 \cite{Butenschoen:2011yh,Butenschoen:2022orc,Ma:2010vd}, with
gaps small against the error bands themselves on logarithmic scale.
This suggests that the perturbative expansions are well behaved.

We are now in a position to assess LDME sets 1--4
with regard to their ability to usefully describe the ATLAS data
\cite{Aad:2014kba,Aaboud:2019wfr,Aad:2014rua} at NLO in NRQCD.
We immediately observe that LDME sets 2 \cite{Gong:2012ug} and 3
\cite{Bodwin:2015iua} lead to negative direct $J/\psi+W/Z$
production cross sections, which is physically unacceptable.
They are only rescued into the positive by the feed-down contributions.
Next, we observe that LDME sets 1--3 (plus the ones
of Refs.~\cite{Chao:2012iv,Han:2014jya,Zhang:2014ybe}, for which we refrain
from showing results for lack of space) lead to predictions that throughout
undershoot the data by about one order of magnitude.
To attribute such a sizable gap to underestimated DPS contributions would
require the cross sections to be overwhelmingly dominated by DPS, in contrast
to the $J/\psi$--$W/Z$ azimuthal-angle analyses of
Refs.~\cite{Aad:2014kba,Aaboud:2019wfr,Aad:2014rua}, which all support SPS
dominance.
This renders LDME set~1 unfavorable, albeit not invalid. 
On the other hand, LDME set~4 \cite{Brambilla:2022rjd} leads to an
underestimation of the data by only a factor of about three, with experimental
and theoretical uncertainties typically touching or overlapping.
We note in passing that this and the other LDME set determined in
Ref.~\cite{Brambilla:2022rjd} have, however, their own problems in
applications beyond the scope of this paper, including negative NLO predictions
for the LHCb measurement of prompt $\eta_c$ production \cite{LHCb:2014oii} and
overshoot of HERA photoproduction data by one order of magnitude.
Furthermore, they involve a delicate fine tuning of negative ${}^3P_J^{[8]}$ and
positive ${}^3S_1^{[8]}$ $J/\psi$ hadroproduction channels canceling to
around 90\%.

To summarize, we have presented the first complete NLO NRQCD predictions of
prompt-$J/\psi$ plus $W/Z$ associated hadroproduction,
tackling $P$-wave loop contributions with an additional large mass scale.
Requiring consistency with ATLAS data
\cite{Aad:2014kba,Aaboud:2019wfr,Aad:2014rua} provides valuable new information
on the interplay of the $J/\psi$, $\chi_{cJ}$, and $\psi(2S)$ LDMEs,
orthogonal to the one encoded in one-particle-inclusive charmonium production
data previously fitted to
\cite{Butenschoen:2011yh,Butenschoen:2022orc,Ma:2010vd,Gong:2012ug,Bodwin:2015iua,Brambilla:2022rjd,Chao:2012iv,Han:2014jya,Zhang:2014ybe},
which has allowed us to critically assess the resulting LDME sets.
While none of the existing LDME sets
\cite{Butenschoen:2011yh,Butenschoen:2022orc,Ma:2010vd,Gong:2012ug,Bodwin:2015iua,Brambilla:2022rjd,Chao:2012iv,Han:2014jya,Zhang:2014ybe}
fully agrees with the world data of prompt $J/\psi$ yield and
polarization when $J/\psi + W/Z$ hadroproduction is included,
Table~\ref{tab:sdcs} will help LDME fitters to find out how far NRQCD
factorization holds at NLO.
The decent description \cite{Brambilla:2022rjd}
of the ATLAS measurement of prompt-$J/\psi$ plus $W$
production \cite{Aaboud:2019wfr,Aad:2014rua}
provides strong evidence for the color octet mechanism
and, once again, exposes the deficiency of the CSM to describe charmonium
production.
Our analysis thus marks an important milestone on the path of scrutinizing
NRQCD factorization.

This work was supported in part by BMBF Grant No.\ 05H18GUCC1 and DFG Grant
No.~KN~365/12-1.

\end{document}